**Remote picometric acoustic sensing via ultrastable laser interferometry**

Yoon-Soo Jang,[1,2,3,*] Dong Il Lee,[1] Jaime Flor Flores,[1] Wenting Wang,[1,4] and Chee Wei Wong,[1*]

[1]Fang Lu Mesoscopic Optics and Quantum Electronics Laboratory, University of California, Los Angeles, CA 90095, USA.

[2]Length and Dimensional Metrology Group, Division of Physical Metrology, Korea Research Institute of Standards and Science (KRISS), 267 Gajeong-ro, Yuseong-gu, Daejeon, 34113, Republic of Korea.

[3]Department of Science of Measurement, University of Science and Technology (UST), Daejeon, 34113, Rep. of Korea.

[4]Communication and Integrated Photonics Laboratory, Xiongan Institute of Innovation, Chinese Academy of Sciences, Xiong'an New Area 071700, China.

* ysj@kirss.ac.kr; cheewei.wong@ucla.edu

**Acoustic detection has many applications across science and technology, from medical to imaging and communications. However, most acoustic sensors have a common limitation in that the detection must be near the acoustic source. Alternatively laser interferometry with picometer-scale motional displacement detection can rapidly and precisely measure sound-induced minute vibrations on remote surfaces. Here we demonstrate the feasibility of sound detection up to 100 kHz at remote sites with ≈ 60 m of optical path length via laser homodyne interferometry. Based on our ultrastable Hz-linewidth laser with $10^{-15}$ fractional stability, our laser interferometer achieves 0.5 pm/Hz$^{1/2}$ displacement sensitivity near 10 kHz, bounded only by laser frequency noise over 10 kHz. Between 140 Hz to 15 kHz, we achieve a homodyne acoustic sensing sensitivity of sub-nm/Pa across our conversational frequency overtones. The minimal sound pressure detectable over 60 m of optical path length is ≈ 2 mPa, with dynamic ranges over 100 dB. With the demonstrated standoff picometric distance metrology, we successfully detected and reconstructed musical scores of normal conversational volumes with high fidelity. The acoustic detection via this precision laser interferometer could be applied to selective area sound sensing for remote acoustic metrology, optomechanical vibrational motion sensing and ultrasensitive optical microphones at the laser frequency noise limits.**

**Introduction**

With length as one of the fundamental physical quantities, ultra-precise length metrology is pervasive across diverse areas in science and technology [1-3]. The current SI definition of the meter is based on the optical path length travelled during 1/299,792,458 of a second in vacuum, with optical metrology serving a pivotal role in ultra-precise dimensional metrology [4,5]. Laser interferometry – in homodyne or heterodyne principles for displacement measurement with sub-wavelength precision – has enabled optical dimensional metrology with advancement of its precision and measurement range [6-10]. However, the precision of interferometric phase-based displacement or distance measurements is practically bounded about several tens of picometer level, from the precision of phase measurements [11-15]. Alternatively, Fabry-Perot (FP) interferometry, which tracks the cavity resonance frequency [16-19], or frequency domain analysis of interferometric signal [20-23] have the capability of picometer level displacement measurement and have been demonstrated for applications of extremely small displacement measurement. Such platforms are widely used to examine gravitational wave searches [1], dynamics on optomechanics [24], membranes [25], and nanomechanical structures [26,27]. In addition to length metrology, various optical and laser sensors have contributed to applications ranging from pressure measurement [28], acoustic sensing [29-32], force and acceleration measurements [33-35], gyroscopes [36], strain sensing [37], earthquake detection [38], to chemical detection [39]. In this study, acoustic sensing can aid in applications such as voice recognition, biological-medical imaging, and ultrasonic sensing.

Through interferometric homodyne stabilization with our Hz-linewidth laser, the 60 m of optical path length is stabilized within a 1σ standard deviation of 2.29 nm, with a measured displacement noise floor of 0.5 pm/Hz$^{1/2}$ near 10 kHz. Examining across the control and error signals in the homodyne metrology up to 100 kHz, the corresponding distance strain noise floor is observed at $1.7 \times 10^{-14}$ ε/Hz$^{1/2}$ with a 1.5 kHz servo bandwidth. Subsequently we demonstrated the remote motional vibrational sensing of a glass beamsplitter under acoustic drives, from 140 Hz to 15 kHz, and quantified the laser homodyne displacement sensitivity as 782.77 nm/V and acoustic sensing sensitivities as sub-nm/Pa across our conversational frequency overtones. The minimal sound level and pressure is determined to be ≈ 40 dB and ≈ 2 mPa, respectively, bounded only by the laser frequency noise. Across the acoustic frequencies, the dynamic range is determined to be between 60 dB to 100 dB, within the laser λ/4 displacement. With the picometric noise metrology and distance stabilization thus demonstrated, we reconstructed real-time sound waveforms and analyzed their frequency spectrograms at the remote 60 m of optical path length, comparing between the control signal and error signal mapping, along with different acoustic overtones. Our demonstrated platform not only allows for remote acoustic sensing in targeted areas but also ultrasound sensing and the adaptation of optical frequency standards towards sound metrology.

**Results**

**Remote picometer displacement measurement using laser homodyne interferometry**

Figure 1(a) shows the concept of our laser homodyne interferometry-based acoustic detection. Sound propagation, when reflected by an interface structure including a window, results in minute vibrations on the window surface. The sound amplitude and frequency information are embedded in window vibrational overtones. With precision laser interferometry, the vibration overtones are retrieved, enabling remote and rapid sensing. Fig. 1(b) depicts the experimental setup for the picometric displacement measurement realization with our precision laser homodyne interferometry. A 1565.54 nm Fabry-Perot (FP) cavity-stabilized ultrastable laser (SLS-INT-1550-100-1, Stable Laser Systems) with 1 Hz linewidth and $10^{-15}$ fractional stability at 1 s is used as the light source in this study. After a single-mode fiber splitter with 9:1 ratio, the shorter arm (10%) is used as the reference while the longer arm (90%; signal) goes through a 40 m fiber delay line which corresponds to an optical path length about 60 m (30 m distance for Michelson-type interferometer) between target and sensing position. A piezoelectric (PZT) actuator, with displacement sensitivity of 2.8 μm/V considering the interferometer roundtrip beam path, controls the optical phase delay line. The laser beam is launched into free-space by a collimating lens and reflected by a beamsplitter window with mm thickness. The reflected beam is collimated into a single-mode fiber and combined with reference arm by a 2×2 optical coupler. The combined reference and signal arm is sent to a balanced photodiode (BPD) to extract the interference signal without dc-offset. The reflected beam is attenuated to ≈ 25 μW to enable a homodyne signal amplitude of ≈ ± 500 mV, such that the displacement sensitivity of homodyne signal is 782.77 nm/V.

The BPD homodyne signal is sent to a proportional-integral (PI) servo controller, with a small fraction of the homodyne signal sent to a FFT analyzer and high-speed oscilloscope for data analysis. The control signal from PI servo controller stabilizes the homodyne signal to zero point for long-term operation and high sensitive sensing via the PZT actuator. Since the control signal range is fixed at ± 10 V, the PZT actuator can compensate a ± 28 μm displacement. A speaker is installed behind the beamsplitter to generate the acoustic input and music. While recording the acoustic frequencies, a passive DC block electrical filter with greater than 1 Hz passband suppresses the long-term drift of optical phase delay due to refractive index change and thermal expansion on the interferometer.

The homodyne detection based optical path stabilization have been widely used in frequency transfer [40,41] and unbalanced arm interferometry [42]. Fig. 2(a) shows the measurements of in-loop optical path stabilization by homodyne detection and PZT actuator over the long delay line, with Figure 2A an example path stabilization over 6.5 ms at 50 kHz update rate. The Gaussian-shaped histogram has a 2.3 nm (1σ) standard deviation. To quantitatively evaluate the optical path length stability, Figure 2A inset shows the resulting Allan deviation. For the short-time scale (20 μs to 0.1 s), the optical path length stability is determined to be 2.3 nm at 20 μs and gradually improves to 40 pm at 0.1 s, with the measurement fitted relation of 12 pm × $\tau_{avg}^{-0.5}$, where $\tau_{avg}$ is averaging time. For longer averaging time more than 0.1 s, the stability remains near 40 pm, equivalent to a time offset of $1.33 \times 10^{-19}$ s (133 zs).

Figure 2(b) shows the displacement amplitude spectral densities obtained from the error and control signals. After the stabilization of the optical path, the error (red line) and control signal (blue line) are plotted together. For comparison, the error signal without the stabilization is plotted in black line, representing the background displacement noise of target window. From Fig. 2(b), we can estimate the PZT servo bandwidth in the optical path stabilization system to be ≈ 1.5 kHz. Inside the servo bandwidth (< 1.5 kHz), displacement amplitude spectral density from the control signal is dominant and its structure is similar to the amplitude spectral density from error signal without stabilization. In other words, the displacement amplitude spectral density from the control signal inside the servo bandwidth is equivalent to the background noise of the interferometer. The error signal inside the servo bandwidth is the residual component of the optical path stabilization. Outside the servo bandwidth (> 1.5 kHz), the displacement amplitude spectral density from the error signal is dominant and its structure is overlaid directly onto the amplitude spectral density of the error signal without stabilization. Above 10 kHz, the displacement amplitude spectral density is bounded by the ultrastable laser and it fluctuates based on the ultrastable laser frequency noise (further detailed in Supplementary Information Section S1) [19, 43]. The lowest background noise attained is at the few pm/Hz$^{1/2}$ level near the Fourier frequency of 10 kHz, based on the frequency noise of the ultrastable 1 Hz laser. As shown in Fig. 2(b), the lowest background noise is estimated to be 0.5 pm/Hz$^{1/2}$, corresponding to a strain of $1.7\times10^{-14}$ ε/Hz$^{1/2}$. By combining the control signal amplitude spectral density inside the servo bandwidth and the error signal amplitude spectral density outside the servo bandwidth, the amplitude spectral density of displacement can be fully reconstructed over a wide range of Fourier frequencies [44].

**Sound sensing via measurement of picometric displacement on the reflectance surface**

Figure 3(a) shows the resulting frequency spectra with input acoustic signals from 140 Hz to 15 kHz by analysis of the error and control signals. Blue and red lines indicate the background noise level from control and error signals respectively, with sensitivities below 10 pm/Hz$^{1/2}$ up to 100 kHz. Control signals for sound detection are used from 140 Hz to 1 kHz and error signals are used from 2 kHz to 15 kHz (the periodic 60 Hz to 300 Hz peaks are from the 60 Hz harmonics of the electrical power line noise). At lower frequency range up to 5 kHz, the background noise is mostly limited by environmental noise since the whole system is not isolated from its surroundings. Note that diagonal dashed lines (light gray, gray and black) indicate theoretically predicted fluctuation of optical path length power spectral density for wind speed of 0.1 m/s, 1 m/s and 10 ms. [45,46] At higher frequency over 10 kHz, the background noise is limited by frequency noise of the ultrastable laser.

Figure 3(b) shows the acoustic-intensity-dependent displacement amplitude spectral densities for 200 Hz, 500 Hz, 2 kHz and 15 kHz, with clearly detected acoustic signatures. The vertical *y*-axis is linear; we use sound level unit of dB, which does not consider weighting factor for human ear response to directly convert sound intensity to sound pressure with units of Pascal. We measure the sound level for all frequency from the speaker using sound level meter with units of dB. We use control signals for 200 Hz and 500 Hz sound detection and error signals for 2 kHz and 15 kHz sound detection. Within intensities from 60 dB to 100 dB for all frequencies, we observe a linear transduction from the input acoustic intensity to the detected intensity in the optical spectrum.

Figure 3(c) shows the summary mapping of the sound intensity (sound pressure) dependent displacement response of the target window from 140 Hz to 15 kHz. The upper horizontal axis is the equivalent sound pressure converted from the sound intensity shown in the lower *x*-axis. Sound intensity is determined by the peak intensity measured in the frequency domain. Sound signals start to appear at 140 Hz due to the relatively high background noise and low output of the driving speaker below this frequency. Detected sound signals of all frequencies have a linear proportion to the input sound pressure and is proportional to the square of sound level in dB units. Across the eight frequencies shown in Fig 3(c), the typical displacement-to-acoustic intensity sensitivity is determined to be sub-nm/Pa. The difference in the sound signal intensity response for each (mechanical and acoustic) frequency arises from the mechanical response transfer function of the target window. In addition, we note that since the PZT used in this study can cover displacement of ± 28 μm, our system detects much louder sound within the servo bandwidth frequency range. Considering the background noise level marked with the open circle, possible sound level measurement range for all frequencies is plotted with the dashed line. For the higher frequency (2 kHz to 15 kHz) where the error signal is used, the minimum detectable sound level is ≈ 40 dB, corresponding to a sound pressure of ≈ 2 mPa. In this range, we define a maximum measurable range as less than the ±$\lambda$/4 vibration level with a marginal safety coefficient of 2, corresponding to an ≈ 100 dB dynamic range in our measurements. For the lower frequency (140 Hz to 1 kHz) where the control signal is used, the minimum detectable sound level varies from 40 dB to 65 dB, equivalent from 2 mPa to 36 mPa. In this range, the maximum measurable range for lower frequency is estimated to be much higher than the case for higher frequencies, since the optical delay line range of the PZT actuator is ± 28 μm these lower frequencies, a dynamic range up to ≈ 100 dB is estimated. We further note that vibration amplitudes larger than $\lambda$/2 can drop the locking state and hence a tighter level of locking is required to measure a large sound when the control signal is used, with a less definite pinpoint of the maximum measurable range for the lower frequencies. Since a sound level of 60 dB is typical for conversations, our scheme is sufficient to detect human voices over remote window. If our system is operated with shorter than 60 m of optical path, the detectable sound level would be lower than 60 dB since the background noise is decreased.

With the Hz-level laser metrology in-place, subsequently we record and reconstructed several music pieces in the laboratory environment using a high-speed 16-bit oscilloscope to show the feasibility of remote and covert sound detection. Fig. 4(a) shows the real-time sound detection of "UCLA fight song" and its spectrogram, with Fig. 4(a) the original sound waveform and Fig. 4(b) its spectrogram over 10 s. Fig. 4(b) and 4(c) show the recorded and reconstructed waveforms from the control (blue) and error (pink) signals respectively. Slow-varying drift of control signal is suppressed by RF high-pass filter (EF599, Thorlabs) with a 400 Hz cutoff frequency. Even though the error signal is locked to the zero point, the unsuppressed components have sound signal information. Since the control signal has more information below 1.5 kHz where most human voice is distributed, the control signal based sound information is clearer than the error signal based information. However, error signal based sound signal includes higher frequency overtones than control signal based one, from the impulse response measurements noted in Supplementary Information Sections S2. As shown in Supplementary multimedia 1, the reconstructed lyrics of "UCLA fight song" is clearly audible for both control and error signal based music records. In frequency domain, error signal based sound signal shows relatively clearer sound signal over the locking bandwidth, as described in previous section. A comparison of our mm-thickness beamsplitter with a few micrometer-thickness pellicle beam splitter is also noted in Supplementary Information Sections S3. Sound signal higher than 5 kHz is attenuated since it is above the mechanical transfer function of the target window but it is still sufficient to receive and distinguish male and female voices remotely as illustrated in supplementary multimedia 1.

Figure 5 shows further examples of the real-time music recording waveform reconstructions and its corresponding spectrograms. Left panels ((a),(c),(e)) are control signal based results and right panels ((b),(d),(f)) are error signal based results. Fig. 5(a)-(d) are songs from a female singer, and Fig. 5(e)-(f) are from a male singer. All data are converted into ".wav" file format (see supplementary multimedia) and these .wav files are converted into spectrograms as shown in Fig. 5 (b),(d),(e). As described in main text, the control signal based waveforms have a stronger signal than error signal based waveforms, while the error signal based waveforms have higher frequency components. From these results, we confirm that our metrology can record both male and female voice overtones at the remote site.

**Discussion**

In summary we have shown remote and local sound detection via picometric homodyne laser interferometry. A Fabry-Perot cavity stabilized Hz-level linewidth laser with $10^{-15}$ fractional frequency instability enables picometric displacement measurement over $\approx$ 60 m of optical path length. Our precision homodyne laser interferometer achieves displacement noise background of 1.5 pm/Hz$^{1/2}$ near 10 kHz, limited by laser frequency noise. We show the measurement capability of sound detection up to 100 kHz at remote locations about 60 m away, with a measurement range extended using high-speed electronics. The measurement method demonstrated in this study shows long-term operation via stabilization of the homodyne signal regardless of phase wrapping by long-term drift. We measure sounds from 140 Hz to 15 kHz to verify frequency-dependent displacement intensities, with acoustic sensing sensitivities as sub-nm/Pa across our conversational frequency overtones. We confirm that our methodology is able to measure sounds ranging from 2 mPa to 2 kPa, with a dynamic range determined between 60 dB to 100 dB, within the laser $\lambda/4$ displacement. With the noise floors and sensitivities determined, we successfully recorded and recreated several music sounds including female and male voices behind a window at typical conversation volumes. We believe that our proposed platform has the potential for laser sound sensing, ultrasound sensing and practical realization of optical frequency standards for acoustic measurements.

**Appendix: Materials and Methods**

**Displacement measurement by control signal analysis.** The fiber stretcher (PZ2, Optiphase) based on piezoelectric actuator has a displacement sensitivity of 5.6 μm/V. In this study, we directly use the output port signal of the servo controller as a control signal, without a voltage amplifier to avoid its voltage noise. Since the output voltage range of control signal is ± 10 V and our interferometer has double path of optical beam line, the fiber stretcher compensates the displacement of ± 28 μm on the target window with sensitivity of 2.8 μm/V. The control signal is directly converted into displacement in a high-speed oscilloscope and FFT analyzer. The control signal is used inside the locking bandwidth of 1.5 kHz.

**Displacement measurement by error signal analysis.** The amplitude of the homodyne signal is fixed to ± 500 mV, equivalent to ± 391.39 nm ($\lambda/4$). We assume that the error signal is linearly proportional to the displacement near zero-point where we use in this study. Displacement sensitivity of the homodyne signal is determined to be 782.77 nm/V. Since the voltage information of error signal is directly converted into displacement in time-domain, the displacement information can be rapidly recorded by a high-speed oscilloscope and FFT analyzer. Since the error signal below the locking bandwidth is suppressed by servo control mechanism, the outside-locking bandwidth of the error signal is valid to detect the acoustic signatures.

27. M. J. Yap, J. Cripe, G. L. Mansell, T. G. McRae, R. L. Ward, B. J. J. Slagmolen, P. Heu, D. Follman, G. D. Cole, T. Corbitt, and D. E. McClelland, "Broadband reduction of quantum radiation pressure noise via squeezed light injection," *Nat. Photon.* **14**, 19-23 (2020).

28. P. F. Egan, J. A. Stone, J. E. Ricker, J. H. Hendricks, and G. F. Strouse, "Cell-based refractometer for pascal realization," *Opt. Lett.* **42**, 2944-2947 (2017).

29. M. Campbell, J. A. Cosgrove, C. A. Greated, S. Jack, and D. Rockliff, "Review of LDA and PIV applied to the measurement of sound and acoustic streaming," *Opt. Laser Technol.* **32**, 629-639 (2000).

30. J. A. Guggenheim, J. Li, T. J. Aleen, R. J. Colchester, S. Noimark, O. Oqunlade, I. P. Parkin, I. Papakonstantinou, A. E. Desjardins, E. Z. Zhang, and P. C. Beard, "Ultrasensitive plano-concave optical microresonators for ultrasound sensing," *Nat. Photon.* **11**, 714-179 (2017).

31. G. Wissmeyer, M. A. Pleitez, A. Rosenthal, and V. Ntziachristos, "Looking at sound: optoacoustics with all-optical ultrasound detection," *Light Sci. Appl.* **7**, 53 (2018).

32. S. Basiri-Esfahani, A. Armin, S. Forstner, and W. P. Bowen, "Precision ultrasound sensing on a chip," *Nat. Comm.* **10**, 132 (2019).

33. E. Gavartin, P. Verlot, and T. J. Kippenberg, "A hybrid on-chip optomechanical transducer for ultrasensitive force measurements," *Nat. Nanotech.* **7**, 509-514 (2012).

34. A. G. Krause, M. Winger, T. D. Blasius, Q. Lin, and O. Painter, "A high-resolution microchip optomechanical accelerometer," *Nat. Photon.* **6**, 768-772 (2012).

35. Y. Huang, J. G. Flor Flores, Y. Li, W. Wang, D. Wang, N. Goldberg, J. Zheng, M. Yu, M. Lu, M. Kutzer, D. Rogers, D.-L. Kwong, L. Churchill, and C. W. Wong, "A chip-scale oscillation-mode optomechanical inertial sensor near the thermodynamical limits," *Laser Photon. Rev.* **14**, 1800329 (2020).

36. Y.-H. Lai, Y.-K. Lu, M.-G. Suh, Z. Yuan, and K. Vahala, "Observation of the exceptional-point- enhanced Sagnac effect," *Nature* **576**, 65-69 (2019).

37. Y. Na, C.-G. Jeon, C. Ahn, M. Hyun, D. Kwon, J. Shin, and J. Kim, "Ultrafast, sub-nanometre-precision and multifunctional time-of-flight detection," *Nat. Photon.* **14**, 355-360 (2020).

38. G. Marra, D. M. Fariweather, V. Kamalov, P. Gaynor, M. Cantono, S. Mulholland, B. Baptie, J. C. Castellanos, G. Vagenas, J.-O. Gaudron, J. Kronjager, I. R. Hill, M. Schioppo, I. Barbeito Edreira, K. A. Burrows, C. Clivati, D. Calonico, and A. Curtis, "Optical interferometry-based

**Figures and Tables**

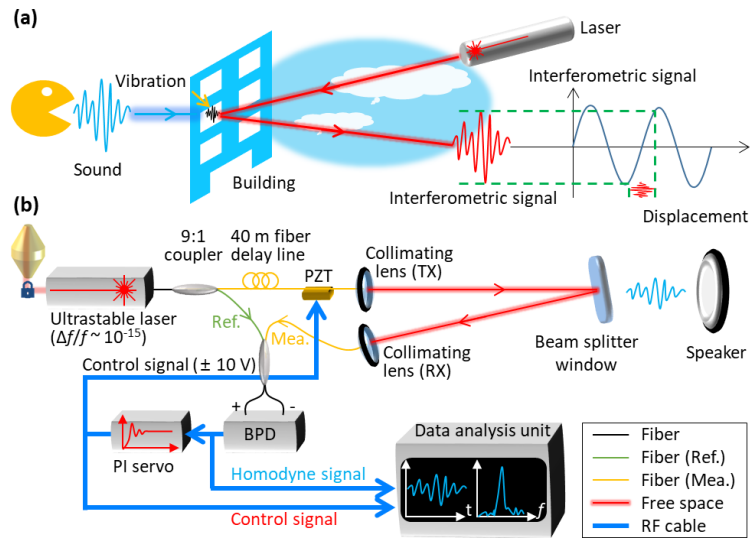

**Figure 1 | Schematic and setup of picometric homodyne laser interferometry based remote acoustic detection.** (**a**) Laser interferometer detects acoustic information engraved on the picometric vibration of the window. The interferometric signal is converted into waveform of sound in the time-domain. (**b**) Measurement interferometer setup with ultrastable few-Hz linewidth laser. BPD: balanced photodetection; PZT: piezoelectric transducer; TX/RX: transmitter and receiver; PI: proportional and integral servo control.

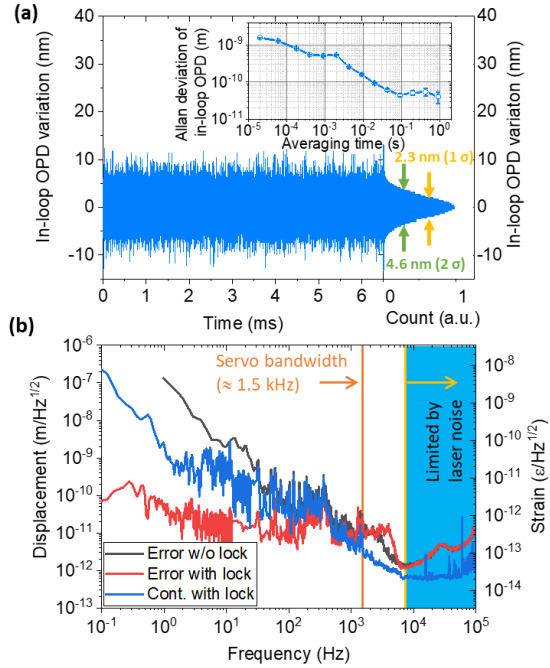

**Figure 2 | Measured displacement amplitude spectral density through optical path stabilization using interferometric homodyne signal.** (a) In-loop measurement of optical path after interferometric homodyne stabilization, sampled over 6 ms and plotted in the time-domain. Right panel: histogram of the stabilized optical path shows Gaussian distribution with 1σ standard deviation of 2.3 nm. Inset: measurement stability verification through Allan deviation. (b) Left axis is the measured displacement amplitude spectral density from error signal (red) and control signal (blue) of the interferometric homodyne stabilization. The laser homodyne displacement noise floor is determined to be 0.5 pm/Hz$^{1/2}$ over the 60 m of optical path length, near 10 kHz. Right axis indicate the corresponding strain, with a noise floor of $1.7\times10^{-14}$ ε/Hz$^{1/2}$. For comparison, free-running homodyne signal without locked optical path stabilization is overlaid in the dark gray plot. Outside the ≈ 1.5 kHz servo bandwidth, displacement from error signal is dominant while displacement from control signal is dominant inside the servo bandwidth.

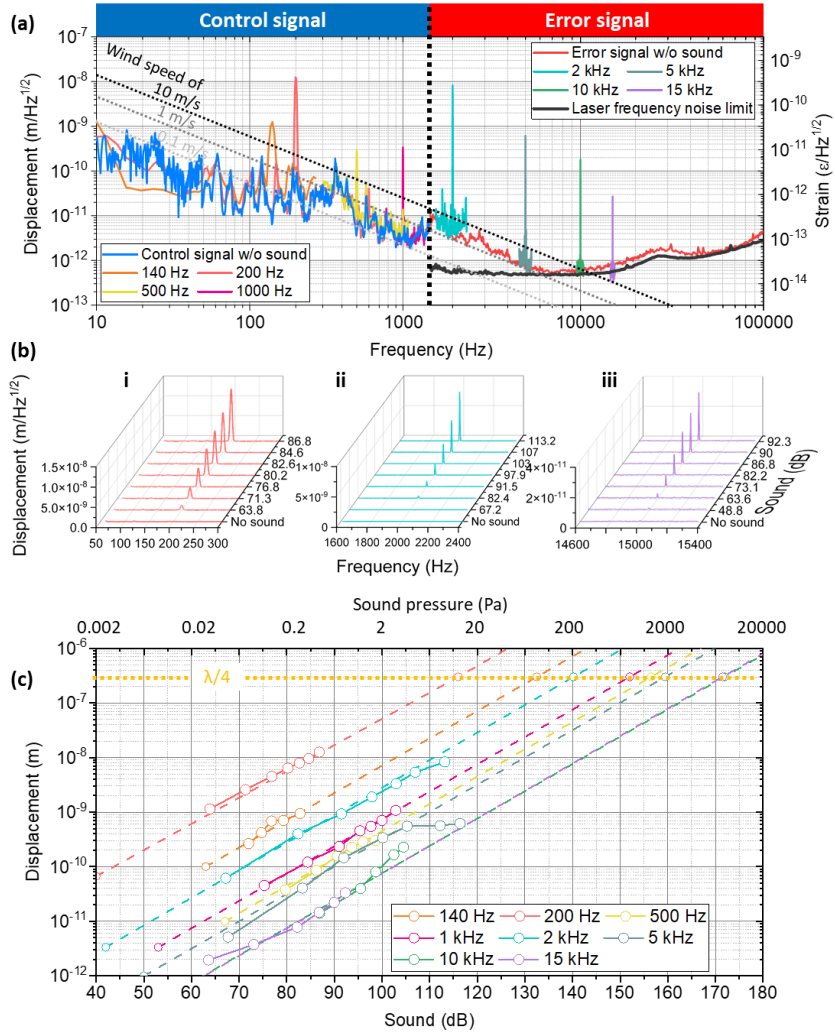

**Figure 3 | Frequency spectra of acoustic sensing for 140 Hz to 15 kHz input signals and varied sound levels, along with mechanical displacement responses.** (**a**) Spectral densities of laser homodyne control and error signal in frequency domain. Blue and red lines indicate background displacement noise measured by the control and error signals respectively. A minimum background displacement noise is found to be ≈ 0.5 pm/Hz$^{1/2}$ near 10 kHz. Dashed lines indicate theoretical predictions of optical path length power spectral density for wind speeds of 0.1 m/s, 1 m/s and 10 m/s. [45,46] (**b**) Intensity-dependent displacement amplitude spectral density displacement in linear scale. Sound level is first measured by a commercial sound meter with unit of dBA. This is then converted into decibel units for direct conversion to sound pressure. For (i) 200 Hz, displacements are measured by the control signal. For (ii) 2 kHz and (iii) 15 kHz, displacements are measured by the error signal. (**c**) Summary displacement response of the target window. Measurement results show from 140 Hz to 15 kHz as a function of sound level (dB) and

sound pressure (Pa). Dashed lines indicate linear fitted lines and open circles are estimated minimum and maximum measurable ranges for each frequency.

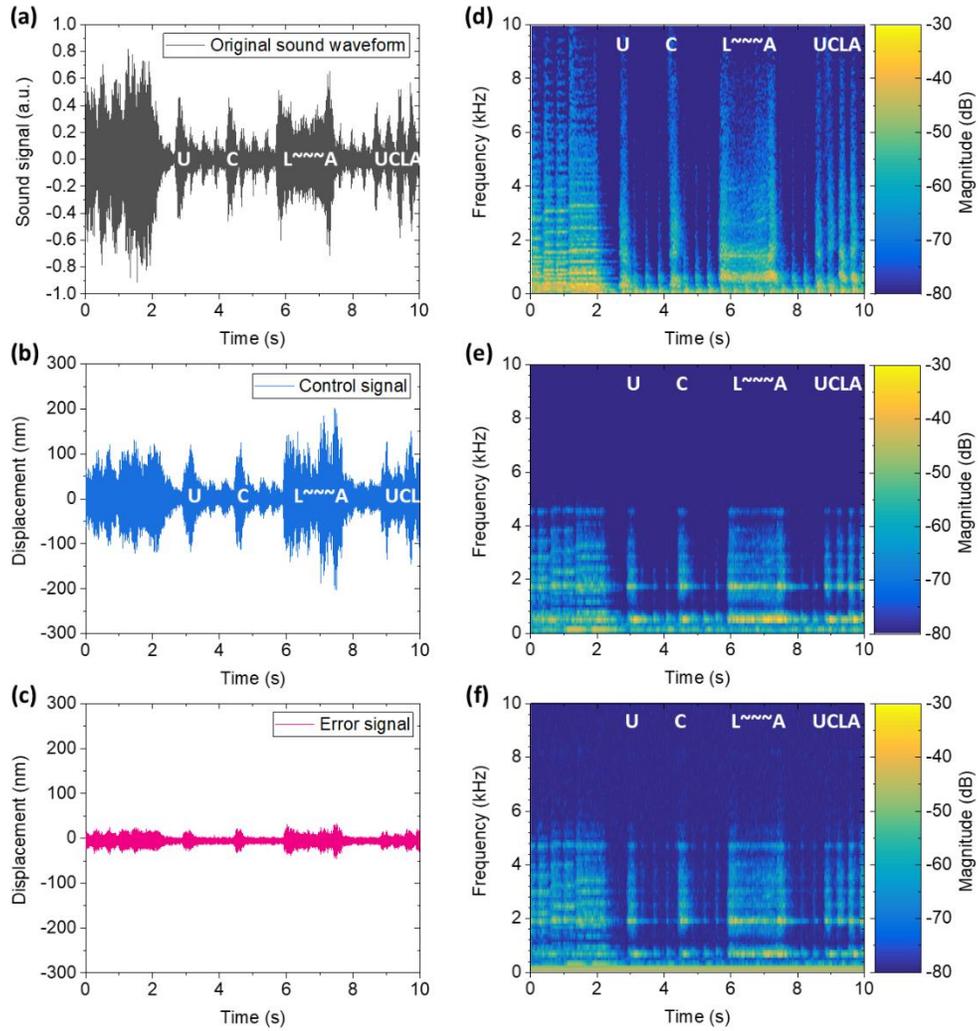

**Figure 4 | Real-time sound reconstruction of "UCLA fight song" and corresponding spectrograms. (a)-(c)** Time-domain waveforms of the original sound (gray), control signal based reconstruction measurement (blue) and error signal based reconstruction measurement (pink). **(d)-(f)** Spectrograms corresponding to **(a)-(c)**, respectively. The control signal based reconstruction has larger signal and higher signal-to-ratio than the error signal based reconstruction. In contrast, the error signal based reconstruction has higher frequency components.

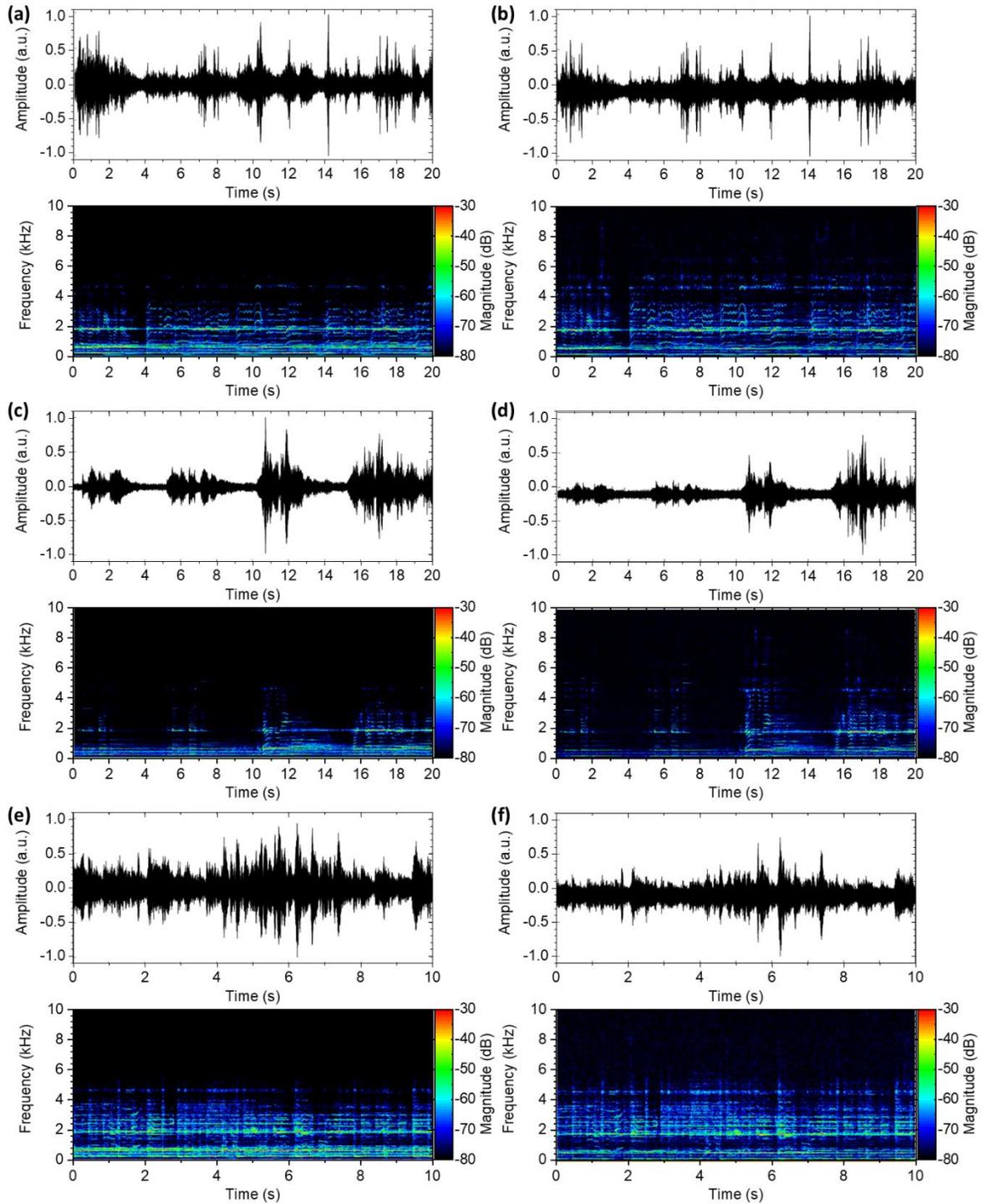

**Figure 5 | Comparison of real-time music recording reconstructions and across different acoustic overtones.** (**a**) Reconstructed control signal waveform and spectrogram of "Shallow1.wav". (**b**) Reconstructed error signal waveform and spectrogram of "Shallow1.wav", with lower amplitude but with higher frequency components distinguished. (**c**) Reconstructed

control signal waveform and spectrogram of "Shallow2.wav". **(d)** Reconstructed error signal waveform and spectrogram of "Shallow2.wav". Likewise a lower amplitude is observed but higher frequency components are distinguished. **(e)** Reconstructed control signal waveform and spectrogram of "HotelCalifornia.wav". **(f)** Reconstructed error signal waveform and spectrogram of "HotelCalifornia.wav", with lower amplitude higher frequency metrology.

**Acknowledgments**: The authors appreciate the helpful discussions with Jinkang Lim, Jiagui Wu and Qingsong Bai. This work is supported by the Office of Naval Research (N00014-16-1-2094), the Lawrence Livermore National Laboratory (contract B622827), and the National Science Foundation. Y.-S. Jang acknowledges support from KRISS (24011043).

**Author contributions:** Y.-S. J. designed and led the work. Y.-S. J. and W.W performed displacement measurement. Y.-S. J., D. I. L. J. F.F. and C. W. W. performed the measured data analysis. All authors discussed the results. Y.-S. J. and C.W.W. prepared the manuscript. C.W.W. supported this research.

**Competing interests:** The authors declare that they have no competing interests.

**Data and materials availability:** All data needed to evaluate the conclusions in the paper are present in the paper and/or the Supplementary Information. Additional data related to this paper may be requested from the authors.

# Supplementary Materials for

**Remote picometric acoustic sensing via ultrastable laser interferometry**

Yoon-Soo Jang,[1,2,3,*] Dong Il Lee,[1] Jaime Flor Flores,[1] Wenting Wang,[1,4] and Chee Wei Wong,[1*]

[1]Fang Lu Mesoscopic Optics and Quantum Electronics Laboratory, University of California, Los Angeles, CA 90095, USA.

[2]Length and Dimensional Metrology Group, Division of Physical Metrology, Korea Research Institute of Standards and Science (KRISS), 267 Gajeong-ro, Yuseong-gu, Daejeon, 34113, Republic of Korea.

[3]Department of Science of Measurement, University of Science and Technology (UST), Daejeon, 34113, Rep. of Korea.

[4]Communication and Integrated Photonics Laboratory, Xiongan Institute of Innovation, Chinese Academy of Sciences, Xiong'an New Area 071700, China.

* ysj@kirss.ac.kr; cheewei.wong@ucla.edu

**This Supplementary Materials consists of the following sections:**
Section S1. Effect of laser frequency noise on displacement noise
Section S2. Real-time and long-distance sound and music recording and re-creation
Section S3. Sound detection with pellicle beam splitter

**Section S1. Effect of laser frequency noise on displacement noise**

An interferometric phase ($\theta$) of Michelson interferometry generated by distance ($L$) can be expressed by $\theta = 2\pi \cdot f_{laser} \cdot (2n_{air}L)/c_o$, where $f_{laser}$ is the laser frequency, $n_{air}$ is the refractive index of air, $c_o$ is speed of light in vacuum, and "2" before $L$ denote round-trip optical beam path. Its first-order partial derivatives are then derived as $\Delta\theta = 2\pi \cdot f_{laser} \cdot (2n_{air}\Delta L)/c_o + 2\pi \cdot \Delta f_{laser} \cdot (2n_{air}L)/c_o$. The first term on the right side is the variation of the interferometric phase by displacement ($\Delta L$) and the second term on the right side is attributed to fluctuations of the frequency noise ($\Delta f_{laser}$). In our study, the frequency noise is not negligible in the phase noise, since we inserted a 40 m fiber delay line to simulate long distance metrology. To verify how laser frequency noise affects the phase noise, the homodyne error signals before and after the stabilization are measured for the Fabry-Perot cavity stabilized ultrastable laser with 1-Hz linewidth and a laser diode with few tens of kHz linewidth. Before the optical path stabilization, the homodyne error signals of ultrastable laser (black) and laser diode (green) are well-matched from 1 Hz to 10 Hz, where environmental drift is dominant. Above 10 Hz, the homodyne error signal of the laser diode is larger than the ultrastable laser. Although a commercially available laser diode is enough to measure picometer displacement at short ranges [S1-S5], the ultrastable laser with few Hertz-level linewidth is necessary for long range picometer displacement measurement [S6-S8].

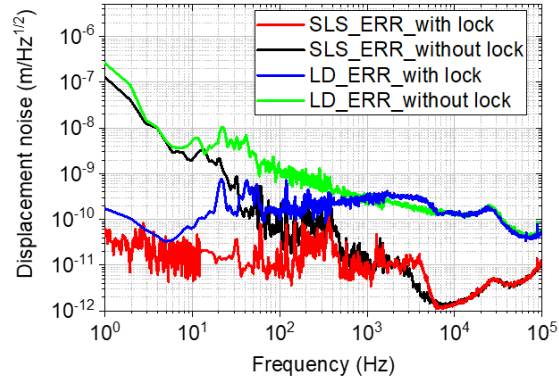

**FIG. S1 | Homodyne error signals before and after optical path stabilization.** Black and red lines are error signals of the ultra-stable laser before and after optical path stabilization, respectively. Blue and green lines are error signals of the laser diode before and after optical path stabilization, respectively.

**Section S2. Real-time and long-distance music recording and re-creation**

As shown in Figures 3 and 4 of the main text, our proposed scheme shows feasibility of sound detection at long distance. Figure S2 shows more examples of real-time sound sensing with known acoustic frequencies of 500 Hz and 5 kHz. The control signal is used to measure 500 Hz sounds and the error signal is used to measure 5 kHz sounds. The red lines indicate low-pass filtered data with cutoff frequencies of 1 kHz and 10 kHz respectively. As shown in Figure S2, the low-pass filtered data show clear sinusoidal signals of both control and error signals.

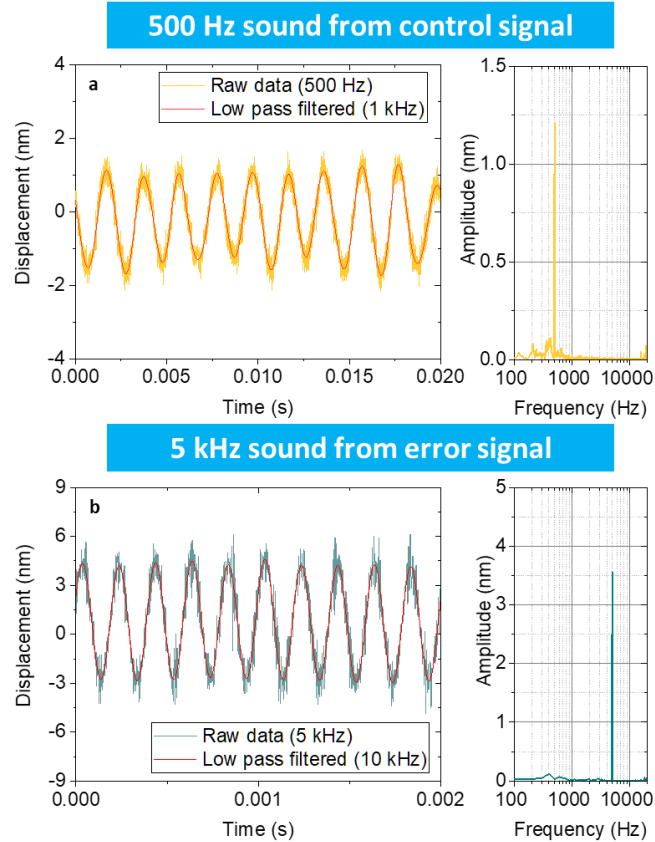

**FIG. S2 | Example of real-time acoustic sensing. a,** 500 Hz sounds recorded by control signal. **b,** 5 kHz sounds recorded by error signal. Sounds are generated by a portable speaker.

Figure S3 shows impulse responses in the time-domain and its spectrograms. The impulse sound is generated by clapping. The upper panels (**a,c**) are control signal based results and the lower panels (**b,d**) are error signal based results. For the impulse response, the displacement measured by the error signal is larger than that measured by control signal because of the bandwidth limitation on control signal. As shown in Figures S3c and d, the error signal based spectrogram shows higher frequency components.

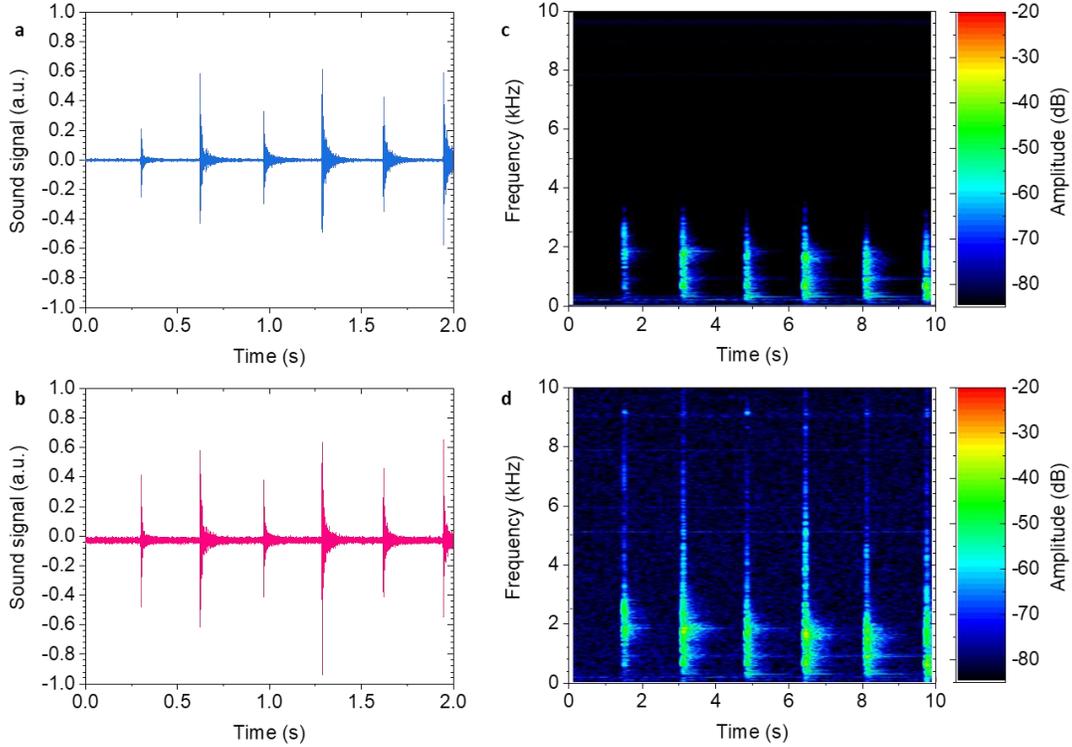

**FIG. S3 | Impulse responses and corresponding spectrograms. a-b,** Time-domain waveforms of control signal based record measurement (blue) and error signal based record measurement (pink). **c-d,** Spectrograms corresponding to **a-b**, respectively. Impulse-like sound is generated by clapping.

**Section S3. Sound detection with pellicle beam splitter**

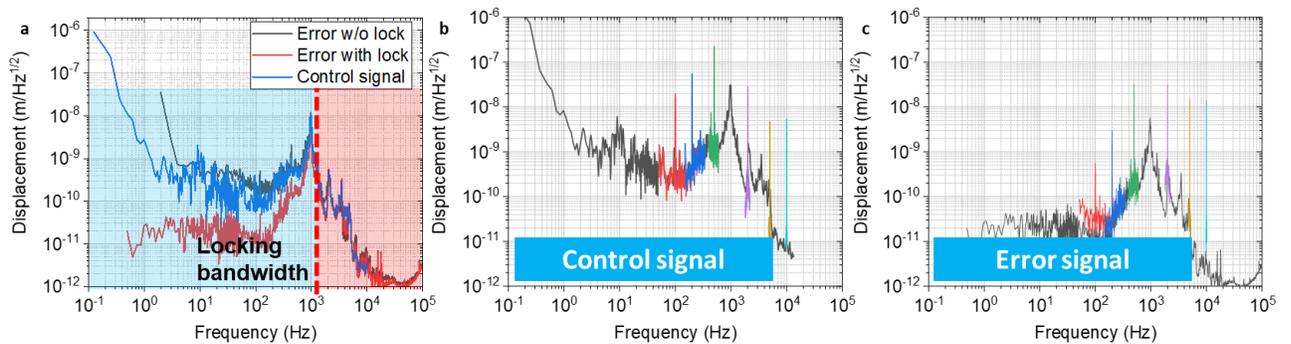

**FIG. S4 | Power spectral density of displacement noise for pellicle beam splitter. a,** Power spectral density of displacement noise from error signal (red) and control signal (blue) with optical path stabilization using interferometric homodyne signal. **b,** Sound sensing using the control signal. **c,** Sound sensing using the error signal.

Since the pellicle beam splitter has few micrometer thickness, its response to sound and environmental noise is more sensitive than the window case described in main text. Figure S4a shows the typical power spectral density of displacement noise floor on the pellicle beam splitter obtained from error and control signals. Error signal (red line) and control signal (blue line) with optical path stabilization are plotted together, and the error signal without the stabilization is also plotted in gray line. Compared to the window case described in main text, the amplitude of displacement is larger over the whole observation frequency. The peak displacement is hundreds of nanometers and observed near 1 kHz, estimated to be mechanical resonance of the pellicle beam splitter. Figures S4b and S4c show example results of frequency-domain sound sensing from 100 Hz to 10 kHz using the control and error signals. Note that near the 1 kHz, the sound signal is not detectable because the amplitude of sound induced vibration is too large to be detected by our system. From 100 Hz to 500 Hz, the control signal based sound signal is larger than the error signal based sound signal. Over the 2 kHz, the error signal based sound signal is larger than the control signal based sound signal.

Figure S5 shows the mapping of sound intensity-dependent displacement response of the pellicle beam splitter from 100 Hz to 15 kHz. All measurement results of displacement amplitude show linear proportionality to the sound intensities. Since thin film-type devices generate large sound induced vibration, our scheme can be used in voice recognition system [S9] and next-generation standard for sound measurement spanning low to high frequencies [S10].

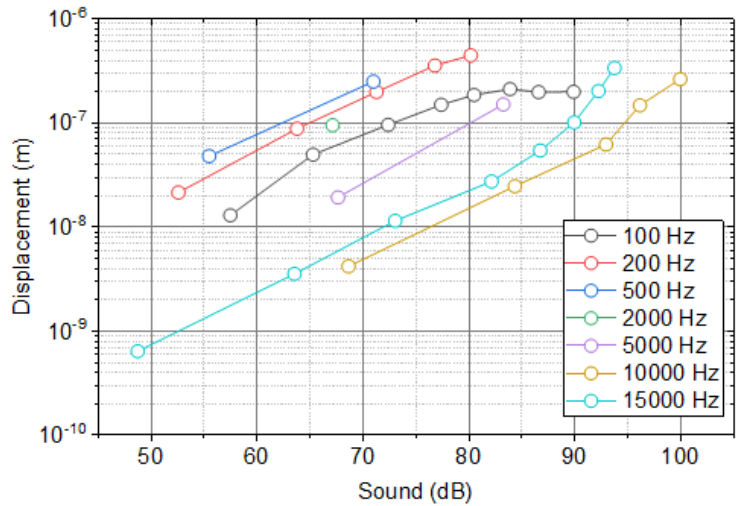

**FIG. S5 | Displacement response of the pellicle beam splitter.** Measurement results show from 100 Hz to 15 kHz as a function of sound level (dB).

**Supplementary References**